\documentclass[conference]{IEEEtran}
\IEEEoverridecommandlockouts

\usepackage{cite}
\usepackage{csquotes}
\usepackage[many]{tcolorbox}
\usepackage{fontawesome5}
\usepackage{booktabs}
\usepackage{float}
\usepackage{orcidlink}
\usepackage{multirow}
\usepackage{url}
\usepackage{hyperref}
\usepackage{comment}
\usepackage{tabularx}
\usepackage{enumitem}
\usepackage{fancybox}
\usepackage{makecell}
\usepackage{csquotes}
\usepackage{colortbl}
\usepackage{mdframed}

\newtcbtheorem{user}{\faWeixin{} \footnotesize Prompt}
{colback=blue!5,colframe=blue!35!black,fonttitle=\footnotesize \bfseries,fontupper=\footnotesize}{th}

\newtcolorbox{colora}{
enhanced,
boxrule=0pt,frame hidden,
borderline west={2pt}{0pt}{gray!50!black},
colback=gray!05!white,
sharp corners
}

\def\BibTeX{{\rm B\kern-.05em{\sc i\kern-.025em b}\kern-.08em
    T\kern-.1667em\lower.7ex\hbox{E}\kern-.125emX}}
\begin{document}

\makeatletter
\newcommand{\linebreakand}{%
  \end{@IEEEauthorhalign}
  \hfill\mbox{}\par
  \mbox{}\hfill\begin{@IEEEauthorhalign}
}
\makeatother


\title{Towards a Framework for Operationalizing the Specification of Trustworthy AI Requirements}


\author{
    \linebreakand
    \IEEEauthorblockN{
        Hugo Villamizar\orcidlink{0000-0003-4142-6967}
    }
    \IEEEauthorblockA{
        \textit{fortiss GmbH} \\
        Munich, Germany \\
        guarinvillamizar@fortiss.org
    }
    \and
    \IEEEauthorblockN{
        Daniel Mendez\orcidlink{0000-0003-0619-6027}
    }
    \IEEEauthorblockA{
        \textit{BTH and fortiss GmbH} \\
        Karlskrona, Sweden and Munich, Germany \\
        daniel.mendez@bth.se
    }
    \and
    \IEEEauthorblockN{
        Marcos Kalinowski\orcidlink{0000-0003-1445-3425}
    }
    \IEEEauthorblockA{
        \textit{PUC-Rio} \\
        Rio de Janeiro, Brazil \\
        kalinowski@inf.puc-rio.br
    }
}

\maketitle

\begin{abstract}
Growing concerns around the trustworthiness of AI-enabled systems highlight the role of requirements engineering (RE) in addressing emergent, context-dependent properties that are difficult to specify without structured approaches. In this short vision paper, we propose the integration of two complementary approaches: \textit{AMDiRE}, an artefact-based approach for RE, and \textit{PerSpecML}, a perspective-based method designed to support the elicitation, analysis, and specification of machine learning (ML)-enabled systems. \textit{AMDiRE} provides a structured, artefact-centric, process-agnostic methodology and templates that promote consistency and traceability in the results; however, it is primarily oriented toward deterministic systems. \textit{PerSpecML}, in turn, introduces multi-perspective guidance to uncover concerns arising from the data-driven and non-deterministic behavior of ML-enabled systems. We envision a pathway to operationalize trustworthiness-related requirements, bridging stakeholder-driven concerns and structured artefact models. We conclude by outlining key research directions and open challenges to be discussed with the RE community.
\end{abstract}


\section{Introduction}
\label{sec:introduction}

Decision-making across a wide range of application areas has become increasingly data-driven and complex. Machine learning (ML) and other artificial intelligence (AI) components are now deeply embedded in software systems that influence critical outcomes—diagnosing diseases, determining creditworthiness, prioritizing emergency responses, and more. As these software systems typically operate with greater autonomy and limited transparency, the decisions they make can have profound impacts on society at large.

In response to growing concerns about fairness, transparency, robustness, and accountability, the concept of trustworthy AI has gained significant attention. Governmental bodies and institutions at, such as the ones in the European Union, have begun to propose regulations for AI trustworthiness (\textit{e.g.}, the EU AI Act ~\cite{EUAIA2024}), while academic and industry experts have published a range of guidelines and ethical principles aimed at fostering responsible AI development~\cite{jobin2019global, floridi2018ai4people, microsoftResponsibleAI}.

Despite this progress, numerous challenges still render the engineering of trustworthy AI-enabled systems cumbersome. These include the difficulty of identifying and specifying socio-technical concerns, dealing with non-deterministic behavior of AI components, and ensuring that requirements align with both stakeholder expectations and regulatory constraints.

Requirements engineering (RE) has increasingly contributed to addressing these challenges, with several studies exploring already how RE practices can support the development of trustworthy AI-enabled systems~\cite{maalej2023tailoring, kwan2021towards, abualhaija2024ai, kosenkov2024regulatory}. As a discipline concerned with the elicitation, analysis, specification, and validation of system properties in context, RE provides structured methods to capture stakeholder needs, quality attributes, and process rules, emphasizing characteristics beyond mere functionality.

However, the operationalization of AI trustworthiness-related requirements remains underexplored. In particular, there is a lack of concrete guidance on how to systematically translate such high-level requirements—technical, social, and regulatory ones—into structured artefacts within RE, let along in a manner that supports seamless engineering. This challenge has also been observed in practice, where practitioners are confronted with a plethora of challenges~\cite{alves2023status} and, thus, express the need for actionable guidelines and knowledge bases to support trustworthiness across the software development lifecycle~\cite{baldassarre2024trustworthy}. Bridging this gap is essential to making trustworthiness practically actionable in real-world system development.

In this vision paper, we explore how two complementary RE approaches—\textit{AMDiRE}~\cite{mendez2015artefact}, an artefact-based approach, and \textit{PerSpecML}~\cite{villamizar2024identifying}, a perspective-based method for specifying requirements in ML-enabled systems—can and should be integrated to support the identification, analysis, and documentation of trustworthy requirements. We position this integration as a step toward enabling concern-aware, artefact-driven RE practices tailored to the unique challenges of trustworthy AI.

\section{Background}
\label{sec:background}

\subsection{Trustworthy AI: Concept and Regulation}
\label{sec:trustworthy-ai}

The concept of \textit{Trustworthy AI} has become central to current discussions about the responsible development and deployment of AI-enabled systems. According to the European Union's High-Level Expert Group on AI (AI-HLEG), an AI-enabled system is deemed trustworthy when it is \textit{“Lawful, complying with all applicable laws and regulations; ethical, ensuring adherence to ethical principles and values; and robust, both from a technical and social perspective, since, even with good intentions, AI systems can cause unintentional harm.”}~\cite{EUHLEG2019}


While the term \textit{Trustworthy AI} has gained prominence in policy and regulatory domains, related terms such as \textit{Ethical AI}, \textit{Responsible AI}, and \textit{Human-Centered AI} are often used in academic and industrial contexts. Although these concepts are sometimes used interchangeably, they differ subtly in focus. \textit{Ethical AI} emphasizes normative values such as fairness and justice; \textit{Responsible AI} emphasizes accountability and governance in the AI lifecycle; and \textit{Human-Centered AI} underscores usability, inclusiveness, and alignment with human needs. These terms can be seen as complementary dimensions or enablers of Trustworthy AI.

To promote trustworthiness in AI-enabled systems, the AI-HLEG\footnote{https://digital-strategy.ec.europa.eu/en/policies/expert-group-ai} identifies seven key requirements:

\begin{enumerate}
    \item \textbf{Human agency and oversight.} Systems should empower human autonomy and support human decision-making, while allowing appropriate oversight mechanisms.
    \item \textbf{Technical robustness and safety.} AI should be secure, reliable, and resilient to errors and adversarial attacks.
    \item \textbf{Privacy and data governance.} Systems must ensure data protection and respect users' privacy rights.
    \item \textbf{Transparency.} Capabilities, limitations, and decision-making processes of AI systems should be transparent and traceable.
    \item \textbf{Diversity, non-discrimination, and fairness.} AI systems should avoid unfair bias and be accessible to all individuals regardless of background.
    \item \textbf{Societal and environmental well-being.} AI systems should benefit all human beings, including future generations, and promote sustainability.
    \item \textbf{Accountability.} Mechanisms should exist to ensure responsibility and auditability across the AI lifecycle.
\end{enumerate}

These requirements establish a high-level normative foundation. However, their translation into concrete, actionable requirements during development remains challenging.

\subsection{AMDiRE}

The \emph{artefact model for domain-independent RE} (short: AMDiRE) emphasizes the specification of structured artefacts, their relationships, and the roles and milestones associated with their development~\cite{mendez2015artefact}. It abstracts from complex processes by defining a set of interrelated artefacts, such as the context specification (\textit{why}), requirements specification (\textit{what}), and system specification (\textit{how}), each with a predefined, detailed content structure. Both aim at supporting seamless engineering while preserving a high degree of flexibility which we deem especially important to RE where the way of working is heavily influenced by various factors. Figure~\ref{fig:amdire_overview} provides an overview of the \textit{AMDiRE} approach.

\begin{figure}
    \centering 
    \includegraphics[width=0.48\textwidth]{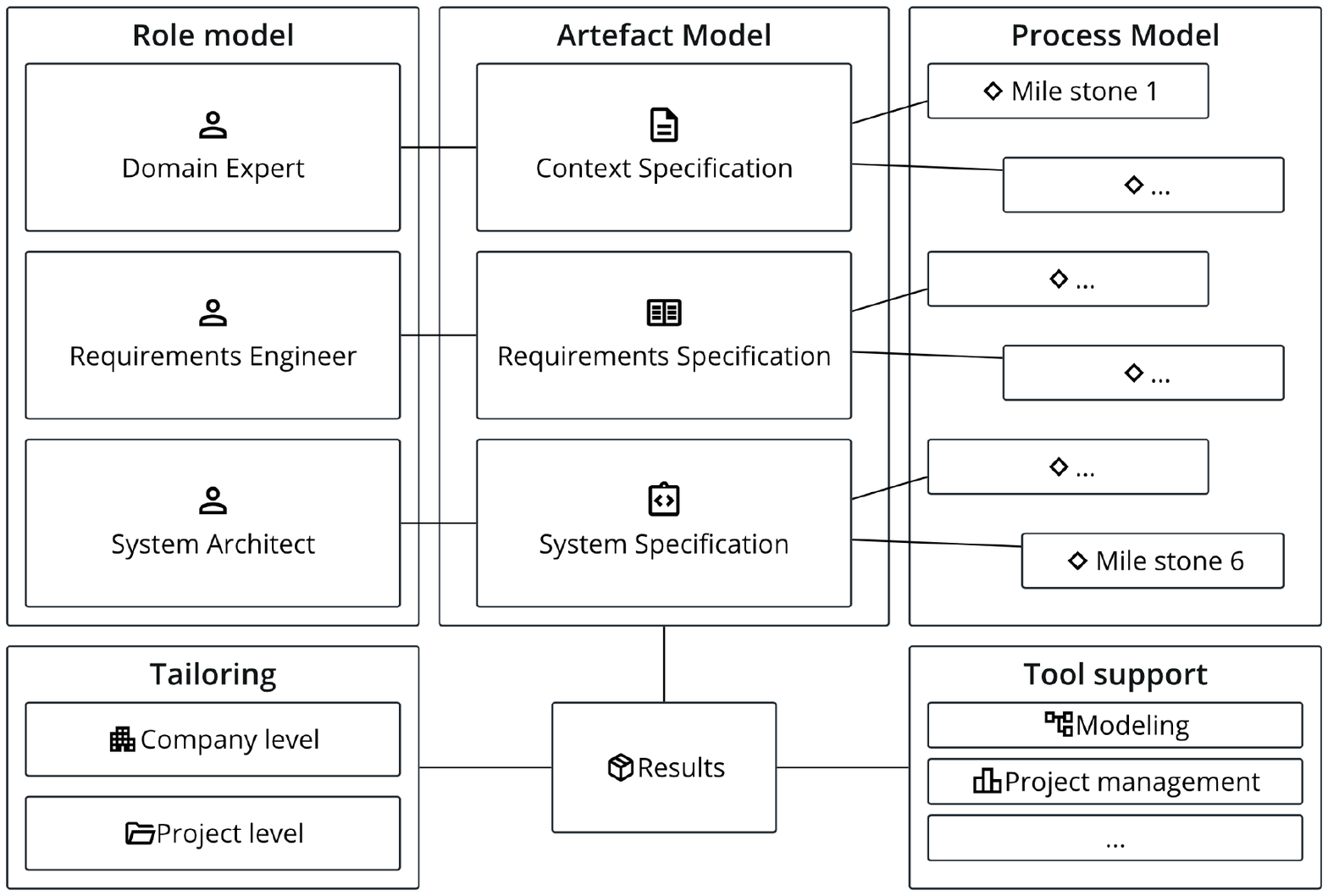}
    \caption{Overview of \textit{AMDiRE}.}
    \label{fig:amdire_overview}
\end{figure}

Its core components—artefact model, role model, and process model—work together to provide a flexible and practical blueprint for specifying requirements across diverse project contexts. The \textbf{role model} defines roles and responsibilities relevant to RE. The \textbf{artefact model} describes a family of artefacts and their interdependencies, specifying what content should be produced and how it is structured. The \textbf{process model} outlines a minimal set of milestones, indicating when artefacts should be created and approved and allows for criteria for their quality assurance.

To support the practical application of these models, \textit{AMDiRE} has been operationalized as a tool\footnote{\label{fn:amdiretool}\url{https://se-toolbox.info/toolbox}}. This tool helps practitioners structure project activities, communicate requirements effectively, and maintain traceability across artefacts throughout the development process.

\subsection{PerSpecML}

In contrast to artefact-centric models, \textit{PerSpecML} introduces a perspective-based approach designed to support the elicitation, analysis, and specification of requirements for ML-enabled systems~\cite{villamizar2024identifying}. It provides structured guidance to analyze 60 concerns related to 28 tasks that practitioners typically face in ML projects, grouping them into five perspectives: system objectives, user experience, infrastructure, model, and data. Together, these perspectives serve to mediate the communication between business owners, domain experts, designers, software and ML engineers, and data scientists. Figure~\ref{fig:perspecml_overview} provides an overview of \textit{PerSpecML}.

\begin{figure}
    \centering 
    \includegraphics[width=0.48\textwidth]{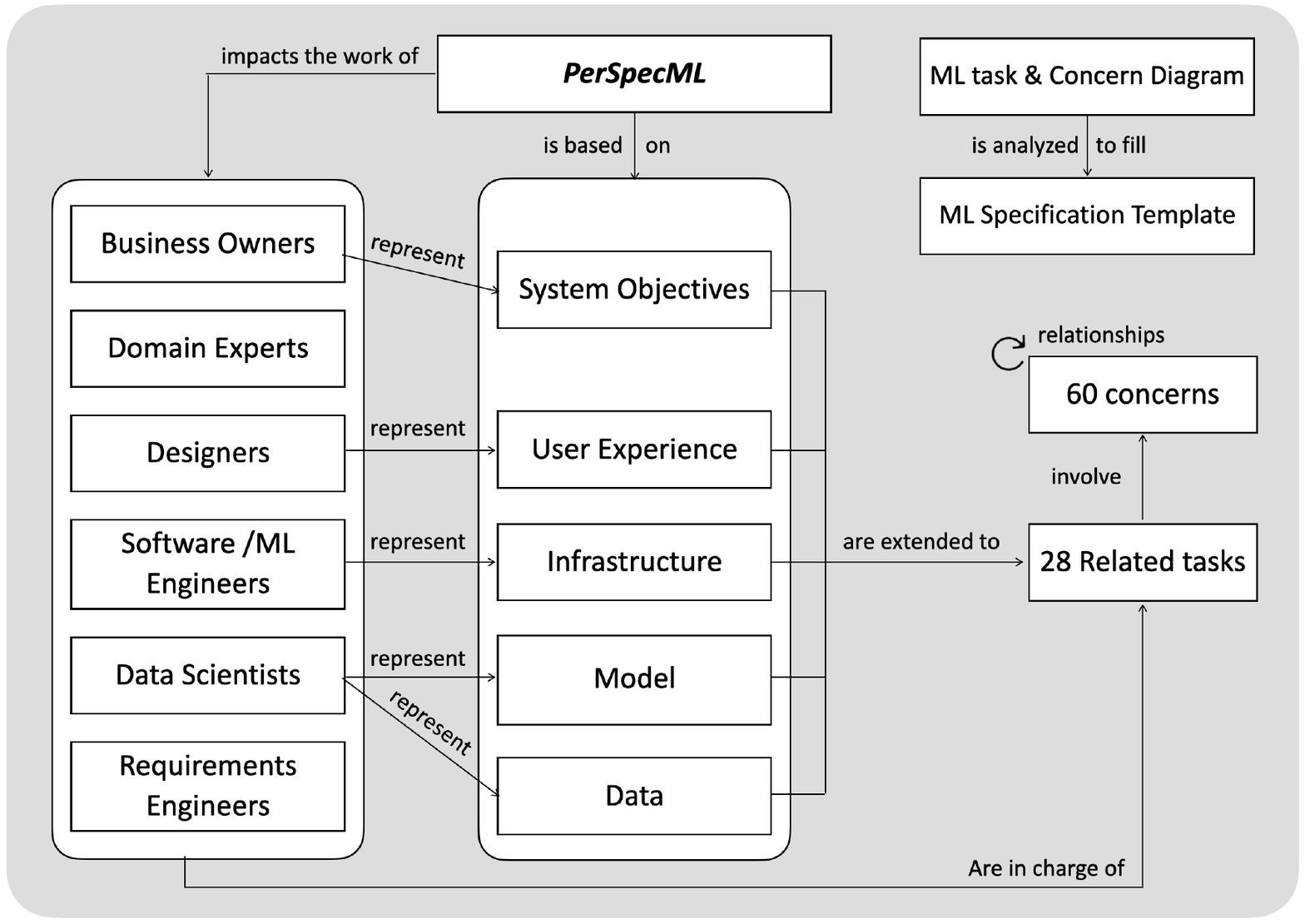}
    \caption{Overview of \textit{PerSpecML}.}
    \label{fig:perspecml_overview}
\end{figure}

At the core of \textit{PerSpecML} is the \textbf{Perspective-based ML Task and Concern Diagram}, which provides a holistic view of the key elements involved in ML-enabled systems. This diagram facilitates early-stage requirements elicitation by enabling stakeholders to describe what the system should accomplish in alignment with project goals. For instance, if a software system requires end-users to understand why a model made a particular decision, the concern of \textit{explainability and interpretability} is triggered. This concern, however, is closely tied to the choice of algorithm—since of them (\textit{e.g}., decision trees, linear regression) are inherently more explainable than others (\textit{e.g.}, deep neural networks, random forests).

In addition to the ML task and concern diagram, \textit{PerSpecML} provides a \textbf{Perspective-based ML Specification Template} to help document and organize requirements derived from the identified concerns. This template aids in translating abstract stakeholder needs into concrete specifications relevant to the ML project.

\section{A Framework Vision for Trustworthy AI Requirements}
\label{sec:vision}


\subsection{Motivation: Complementary Strengths and Limitations}

\textit{AMDiRE} was originally conceived for deterministic software-intensive systems and does not explicitly account for the characteristics of AI-enabled systems. Nonetheless, it offers a robust mechanism to structure the RE process around artefacts such as context, requirements, and system specifications, along with well-defined roles and milestones. These elements remain highly relevant for software systems that incorporate AI components and must address trustworthiness concerns.

\textit{PerSpecML}, on the other hand, was developed with a focus on ML-specific contexts. While it was not designed with trustworthiness as its primary focus, it does address several related concerns—such as accountability, explainability, bias and fairness, and security and privacy. Its strength lies in offering a stakeholder-centric perspective of ML tasks, helping practitioners uncover interrelated concerns across different roles and project stages.

Together, these approaches offer complementary capabilities: \textit{AMDiRE} contributes structured artefact models (specifications) suited at different levels, while \textit{PerSpecML} provides an ML context-aware view of stakeholder concerns. Building on these strengths, our envisioned framework aims to bridge the gap between high-level trustworthiness goals and their concrete manifestation in RE artefacts.

\subsection{Our Vision: From Concerns to Artefact Models}

The idea is to customize and extend \textit{AMDiRE}'s artefact-centric structure to make it suitable for AI-enabled systems, particularly those with trustworthiness concerns, while using the concern- and stakeholder-driven orientation of \textit{PerSpecML} to guide the identification and analysis of relevant requirements. This envisioned framework will be developed through the following steps:

\begin{itemize}
    \item \textbf{Scoping and generalizing beyond ML:} Since \textit{PerSpecML} was initially developed for ML-specific contexts, we will reassess its applicability to the broader AI landscape, including domains such as generative AI. This may require identifying new components that are unique to these emerging paradigms and ensuring they are adequately represented in the framework.

    \item \textbf{Extension of concerns, tasks, and stakeholders:} While \textit{PerSpecML} already incorporates several concerns relevant to trustworthy AI, we aim to extend its concern catalog and associated ML-related tasks by incorporating additional requirements derived from regulatory bodies, such as the EU AI Act~\cite{EUAIA2024}. This includes introducing trustworthiness aspects that are currently underrepresented (\textit{e.g}., societal and environment well-being) and modeling regulatory institutions as first-class stakeholders within the framework.

    \item \textbf{Modeling new relationships and dependencies:} The newly introduced concerns will be analyzed for interdependencies with existing elements (tasks, stakeholders, system goals). These relationships will be modeled to support reasoning about trade-offs and consistency during elicitation and specification.

    \item \textbf{Bridging to artefact models:} Leveraging the artefact-centric structure of \textit{AMDiRE}, the extended concern model will be mapped to corresponding artefacts. This allows for systematic documentation and traceability of trustworthiness-related requirements throughout the development process.  Of special interest will be the synthesis with the already existing artefacts in AMDiRE.

    \item \textbf{Tool-supported operationalization:} We plan to operationalize the framework as a tool that builds upon existing assets from both \textit{AMDiRE} and \textit{PerSpecML}. This tool will support practitioners in identifying, analyzing, and specifying trustworthy AI requirements through guided workflows, templates, and visualization of concerns and artefacts.
\end{itemize}

Figure~\ref{fig:framework-overview} illustrates the envisioned framework. It shows how foundational elements from \textit{AMDiRE}, \textit{PerSpecML}, and regulatory guidance (\textit{e.g}., EU AI Act) are reused and extended across the framework's core modules.

\begin{figure}
    \centering
    \includegraphics[width=\linewidth]{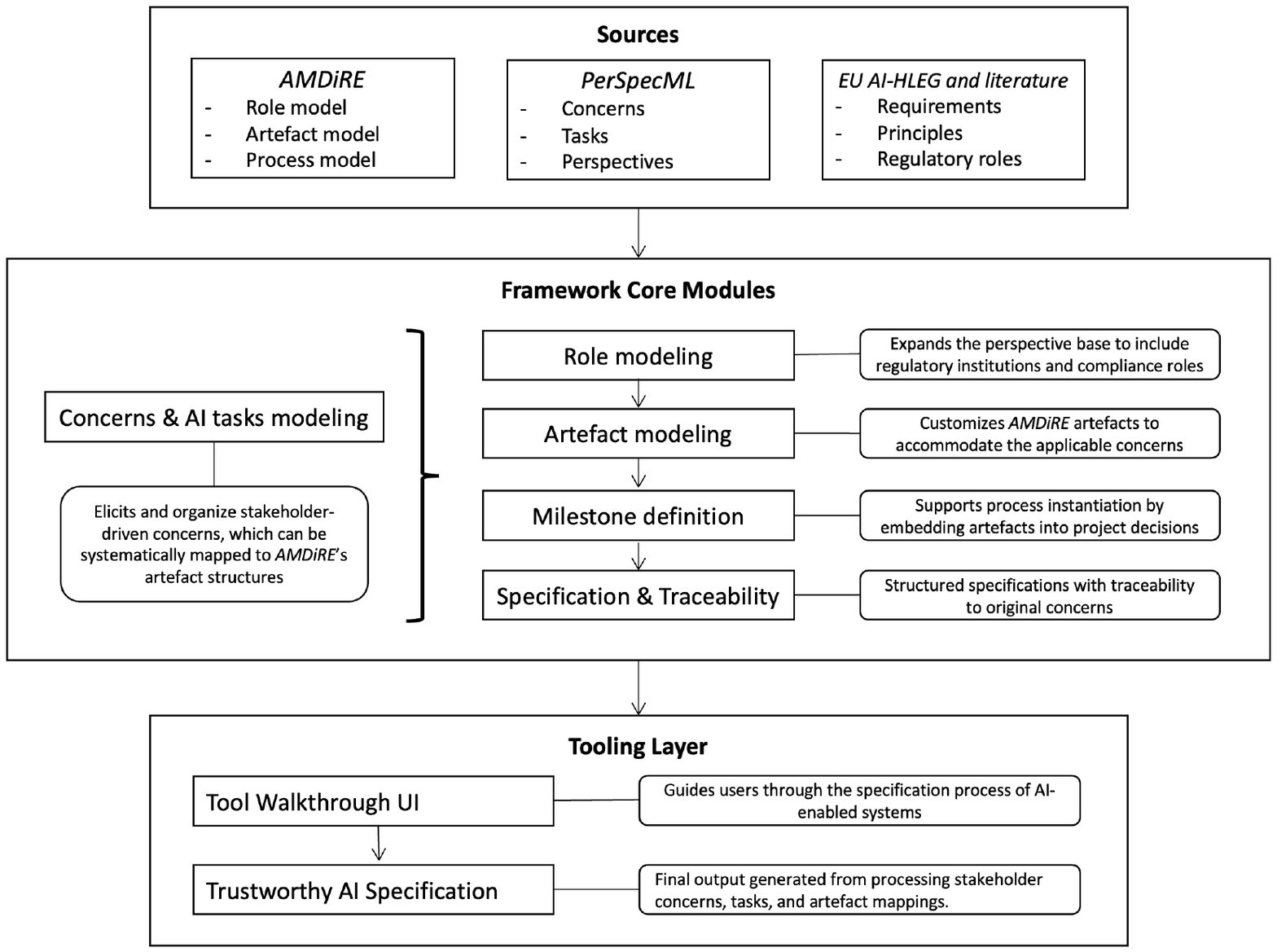}
    \caption{Overview of the envisioned framework.}
    \label{fig:framework-overview}
\end{figure}

\subsection{Illustrative Example}

To demonstrate how the envisioned framework can operate in practice, we present an illustrative example.

\textbf{Context.} A software development team is designing an AI-enabled hiring platform that uses natural language processing (NLP) to assist human resources (HR) departments in screening job applications. The system must comply with trustworthiness requirements.

\textbf{Concerns and AI Tasks Modeling.} Using the extended version of \textit{PerSpecML}, the team engages stakeholders, including HR personnel, data scientists, legal experts, and regulators, to identify relevant tasks, associated concerns, and potential trade-offs. Table~\ref{tab:concerns} summarizes these insights.

\begin{table}[h]
\scriptsize
\centering
\caption{Sample Tasks and Trustworthiness Concerns}
\label{tab:concerns}
\begin{tabularx}{\columnwidth}{|X|X|X|}
\hline
\textbf{Task} & \textbf{Stakeholders} & \textbf{Concerns} \\
\hline
Train NLP model & Data Scientists & Fairness, Bias\\
\hline
Generate model explanations & Data Scientists, Domain Experts & Explainability, Human Oversight \\
\hline
Validate regulatory compliance & Legal Experts, Regulators & Transparency, Traceability, Non-discrimination \\
\hline
\end{tabularx}
\end{table}

\textbf{Mapping to Artefacts.} The concerns and stakeholder perspectives are then mapped to artefacts using \textit{AMDiRE}'s model. Table~\ref{tab:mapping} illustrates this mapping.

\begin{table}[h]
\scriptsize
\centering
\caption{Mapping Concerns to Artefacts}
\label{tab:mapping}
\begin{tabularx}{\columnwidth}{|X|X|X|}
\hline
\textbf{Concern} & \textbf{Mapped Artefact} & \textbf{Example Entry} \\
\hline
Regulatory Compliance & Context Specification & “The system must comply with EU AI Act requirements” \\
\hline
Explainability & System Specification & “Each model decision must be accompanied by a plain-language explanation.” \\
\hline
Model Robustness & System Specification & “The system architecture shall performance under input perturbations.” \\
\hline
Stakeholder Roles & Role Model & Legal expert validates artefacts; HR defines ranking logic. \\
\hline
\end{tabularx}
\end{table}

\textbf{Artefact content instantiation.} Once the concerns are mapped to artefact types, practitioners can use the predefined content structures of each artefact—such as goals and stakeholder models in the \textit{Context Specification}, or usage and quality models in the \textit{Requirements Specification} to guide consistent and traceable trustworthy AI specification. \textit{PerSpecML} can further support this step by providing structured concern descriptions and stakeholder insights to inform artefact content.

\textbf{Tool-Supported Operationalization.} We envision that practitioners can use a tool—extended from the existing \textit{AMDiRE} environment\footref{fn:amdiretool} to support the structured specification of trustworthy AI requirements, including concern navigation, stakeholder input, and artefact generation.
\section{Related Work}
\label{sec:related_work}

A rapid review by Barletta \textit{et al}.~\cite{barletta2023rapid} recently reported that no single framework supports both technical and non-technical stakeholders across all phases of the software development lifecycle. Their findings underscore the absence of comprehensive frameworks that encompass the full range of Responsible AI principles. Focusing more narrowly on SE practices, Baldassarre \textit{et al}.~\cite{baldassarre2024polaris} proposed \textit{POLARIS}, a framework structured around four key principles: explainability, fairness, security, and privacy. While \textit{POLARIS} supports AI developers in aligning system properties with core trustworthiness aspects, it does not offer concrete mechanisms for specifying these requirements or integrating them into structured RE artefacts. From an RE viewpoint, Ahmad \textit{et al}.~\cite{ahmad2023requirements} introduced a layered framework for human-centered AI. Their approach centers on eliciting and visualizing requirements derived from human-centered principles. While closely aligned with our goals, their framework focuses more on usability and human values than on operationalizing structured trustworthiness requirements for AI-enabled systems. These exemplary works shed light on critical aspects of trustworthy AI and contributed valuable perspectives to the field. Our intention is to complement these efforts by building on existing RE methods (\textit{AMDiRE} and \textit{PerSpecML}) to support the structured specification of trustworthiness-related requirements in AI-enabled systems.







\section{Conclusions}
\label{sec:conclusions}

This short paper introduces a vision for a framework that supports the specification of trustworthy AI requirements by drawing on and extending the foundational components of \textit{AMDiRE}~\cite{mendez2015artefact} and \textit{PerSpecML}~\cite{villamizar2024identifying}.

A key direction for future research lies in refining and validating the proposed mappings between trustworthiness concerns and artefact types. This includes understanding how different categories of concerns, such as explainability, human oversight, and robustness, interact with contextual, functional, and system-level specifications. Adapting the concern model to the broader landscape of AI beyond ML raises new methodological and conceptual challenges that must be addressed to ensure relevance and completeness. We therefore postulate that the development can only yield meaningful results if continuously evaluated based in practical settings, \textit{e.g}. through case studies as part of strong industry collaborations. As regulations evolve and concerns continue to grow around the responsible use of AI, the RE community has an unique role to play in shaping methods that are both rigorous and responsive to emerging needs. Our positioning will hopefully contributing to foster this debate further.

\bibliographystyle{IEEEtran}
\bibliography{references}

\end{document}